# Edge Couplers with relaxed Alignment Tolerance for Pick-and-Place Hybrid Integration of III-V Lasers with SOI Waveguides

Sebastian Romero-García, Bahareh Marzban, Florian Merget, Bin Shen, Jeremy Witzens

*Abstract*—**We report on two edge-coupling and power splitting devices for hybrid integration of III-V lasers with sub-micrometric silicon-on-insulator (SOI) waveguides. The proposed devices relax the horizontal alignment tolerances required to achieve high coupling efficiencies and are suitable for passively aligned assembly with pick-and-place tools. Light is coupled to two on-chip single mode SOI waveguides with almost identical power coupling efficiency, but with a varying relative phase accommodating the lateral misalignment between the laser diode and the coupling devices, and is suitable for the implementation of parallel optics transmitters. Experimental characterization with both a lensed fiber and a Fabry-Pérot semiconductor laser diode has been performed. Excess insertion losses (in addition to the 3 dB splitting) taken as the worst case over both waveguides of respectively 2 dB and 3.1 dB, as well as excellent 1 dB horizontal loss misalignment ranges of respectively 2.8 μm and 3.8 μm (worst case over both in-plane axes) have been measured for the two devices. Back-reflections to the laser are below -20 dB for both devices within the 1 dB misalignment range. Devices were fabricated with 193 nm DUV optical lithography and are compatible with mass-manufacturing with mainstream CMOS technology.**

## I. INTRODUCTION

HYBRID integration of prefabricated III-V laser diodes with silicon-on-insulator photonic circuits is the mainstream approach in current industrial practice for overcoming the lack of an efficient electrically-pumped silicon-based light emitter [1,2]. Hybrid integration provides a state-of-the-art reliable and low power consumption light source with low threshold currents, low relative intensity noise and high output power capabilities. Furthermore, the prescreening of the laser before assembly can contribute in reducing yield fallout in mass manufacturing. However, hybrid integration is also associated with some major challenges:

First, a major difficulty arises from mode mismatch. The typical beam dimensions of a semiconductor laser are on the order of 1 μm in the vertical direction and on the order of 1 to a few microns in the lateral (in-plane) direction, while the typical dimensions of single mode high confinement waveguides fabricated in thin SOI (~220 nm) are on the order of hundreds of nanometers. This mode mismatch leads to a low coupling efficiency when both elements are directly coupled. In order to solve this problem, several solutions based on near-vertical coupling with diffraction gratings [3] in combination with a laser micro-package [1,4], as well as edge coupling with spot-size converters (SSCs) [2, 5-8] have been demonstrated. The second and in practice most problematic challenge is the required precision during the assembly of the components. This has led in many cases to the requirement of active alignment, wherein the laser is powered on during assembly and an optical signal is collected as feedback in order to optimize the alignment.

With near-vertical coupling, the laser diode is placed on an external submount where the beam is resized with a ball lens and rotated 90 degrees with a prism or mirror [1,4]. The redirected light is coupled into the photonic chip by means of a planar grating and finally focused or tapered down to the interconnection waveguide width within the plane of the chip. Due to the spot size conversion to an approximately 10 μm spot at the chip interface, a placement tolerance of less than 1 dB excess loss over a 4 μm range have been reported, allowing wafer-scale assembly and mass manufacturing [1,4]. However, this scheme relies on the fabrication of state-of-the-art grating couplers [3] in order to achieve coupling efficiencies higher than 40% (-4 dB), constraining in particular the thickness of the silicon film or the buried oxide (BOX) of the SOI wafer and also requiring the fabrication and assembly of an additional laser submount.

On the other side, edge coupling solutions with spot size converters for thin SOI waveguides usually consist of an inverse taper that gradually reduces the waveguide width far below the single mode condition in the proximity of the chip facet. Such reduction permits an expansion of the waveguide mode profile to a point that matches the size of the laser beam in both the vertical and horizontal directions, leading to high coupling efficiencies [5].

This work was supported by the European Research Council (ERC FP7/2011-2016 No. 279770) and the European Commission's Seventh Framework Programme (CIG FP7/2011-2015 No. 293767).

The authors are with the Integrated Photonics Laboratory (IPH), RWTH Aachen University, Sommerfeldstr. 24, Aachen 52074, Germany (e-mail: sromero@iph.rwth-aachen.de; fmarzban@iph.rwth-aachen.de; fmerget@iph.rwth-aachen.de; bshen@iph.rwth-aachen.de; jwitzens@iph.rwth-aachen.de)

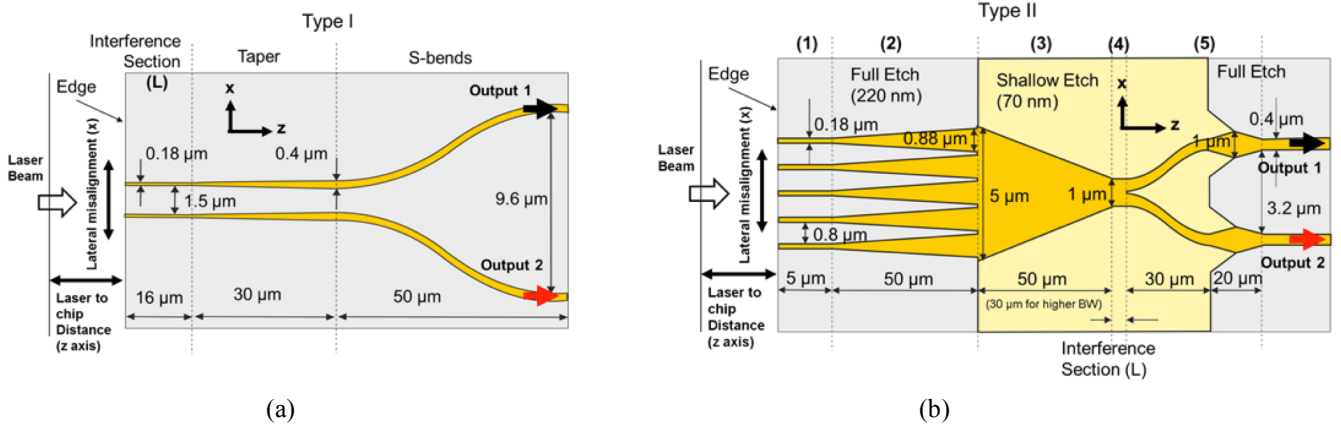

(a)                                                                                          (b)

Fig. 1. Edge coupler schematics: (a) Top-view of the type I edge coupler with two input waveguides (b) Top-view of the type II edge coupler with five input waveguides. The grey areas correspond to a fully etched SOI device layer, the light yellow areas to partially etched regions (150 nm thick slab) and the deep yellow areas to un-etched regions (220 nm thick silicon).

Inverse tapers have additional advantages such as a wide bandwidth, low reflection and low polarization dependent loss. However, conventional inverse tapers exhibit misalignment tolerances of less than ±1 µm, a requirement that is too stringent to be guaranteed by commercially available flip-chip bonders with passive alignment. A relatively modest improvement in alignment tolerance would bring these devices in range of pick-and-place capability [9]. Coupling to a wider waveguide made of a low index polymer or SiON overlay prior to the inverse taper has been proposed to alleviate this problem [5-7], but at the expense of a higher complexity in the fabrication process. Furthermore, the corresponding alignment problem remains single mode (in that light has to be coupled to the ground mode of the overlay waveguide), so that a fundamental trade-off between alignment tolerance and peak insertion efficiency remains in the absence of beam reshaping on the laser side.

Finally, an important metric resides in the amount of chip real estate that has to be allocated purely to the light source. In the case of flip-chip attachment [8,10-13], this area is given directly by the size of the laser, typically on the order of 0.1 mm². Some applications, like optical interconnects with parallel transmission channels [1, 2, 10], make an effective use of the area dedicated to the light source and the high optical power available by distributing the coupled light amongst several parallel transmitters, reducing the final footprint per channel.

In this work, we analyze and experimentally demonstrate how the combination in a compact multimode edge coupler of both optical coupling and equal power splitting into two single mode waveguides can relax the horizontal alignment tolerances required by conventional single mode edge coupling devices. In Section II of this paper, we describe the design of the multimode edge coupling devices and analyze the trade-off between misalignment tolerance and coupling efficiency. In Section III, the experimental characterization of the devices with a lensed fiber and a Fabry-Pérot laser diode is presented. The best device in terms of misalignment tolerance exhibits 3.1 dB excess insertion loss (in addition to the 3 dB splitting losses and measured as the worst case over both waveguides) and a 1 dB horizontal misalignment tolerance of 3.8 µm (measured over both in-plane axes, x and z in Fig. 1). A simpler device relying on the same basic concepts, but with fewer interfaces, exhibits insertion losses of 2 dB and a 1 dB horizontal misalignment tolerance of 2.8 µm.

## II. MULTIMODE EDGE COUPLER DESIGN

### A. Introduction

In conventional edge-coupling structures, the receiving waveguide is designed to be single-mode. High coupling efficiencies can be accomplished when the field distribution of the waveguide mode matches the profile of the light source [14]. This condition can be well satisfied when the center of the light source is perfectly aligned with the center of the coupling waveguide. However, a misalignment between the two elements directly produces a reduction in coupling efficiency. The sensitivity to lateral misalignment can be reduced by broadening the beam in the in-plane dimension on the SOI chip, for example by tapering the waveguide to a width much wider than the width of the laser beam. This however results in a reduced peak insertion efficiency at perfect alignment unless the laser beam is also reshaped, for example with a ball lens in a micro-package. Thus, these two metrics, peak insertion efficiency and alignment tolerance, are traded-off against each other in the case of simple edge coupling. The underlying cause for this trade-off is ultimately a mathematical imperative arising from the reciprocity principle. In a passive system, a degree of freedom of the input beam, here its center position, has to be accommodated by a degree of freedom of the light on the receiving chip in order to maintain high insertion efficiency. In the devices we are presenting, by providing a degree of freedom given by the relative phase between two output waveguides, the trade-off between peak insertion efficiency and alignment tolerance can be overcome.

The top view schematics of the designed structures can be seen in Fig.1. The simpler of the two edge-couplers investigated here (Fig. 1(a)) essentially consists of two coupled waveguides and is referred to as type I in the following. The more complex device transitions between an array of inverse tapers and a multi-mode interference (MMI) section prior to splitting the light into two waveguides (Fig. 1(b)) and is referred to as type II. Both structures are multimode in the horizontal direction. With this coupling scheme, the light source has the potential of exciting not only the fundamental mode, but also an additional higher order mode, while splitting the power from both modes equally between the two output waveguides. A misalignment in the horizontal axis inside a range of several micrometers results in a different excitation of these modes in terms of their amplitude, but not in a significant loss of the total coupled power (see Fig. 2). A properly sized interference region ensures that the power is also equally split between both waveguides.

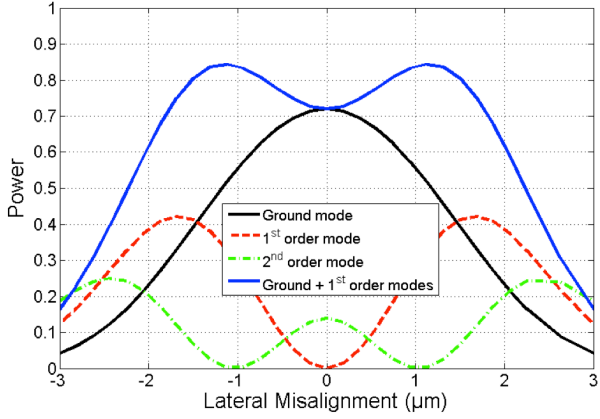

Fig. 2. Power coupled into the first three modes at the input of the type II edge coupler shown in Fig. 1(b) with 5 input waveguides (width = 180 nm, gap = 0.8 μm). It is apparent that the majority of the power is coupled to the first two modes (blue curve) for misalignments within ±2 μm.

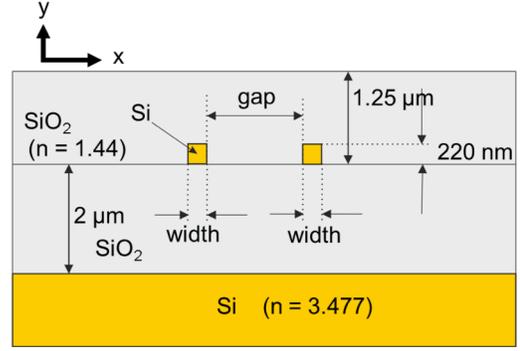

Fig. 3. Cross-section of the type I edge coupler with two input waveguide tips (Fig. 1(a)). Silicon is show in yellow and SiO₂ in grey. Assumed material refractive indices at λ=1550 nm are indicated.

The integration of the splitting function in the same device allows redistributing the coupled power between two single mode waveguides. In a similar fashion to a 3 dB splitter based on a directional coupler, the length is designed to achieve a phase shift of ±π/2 (quadrature condition) between the two lower order modes of the super-structure. The quadrature condition ensures that no matter the relative amplitude of these two modes, equal power is coupled to both output waveguides. While the relative amplitude of the modes is changed as the laser is moved back and forth in the lateral direction along the interface, their relative phase is not (excluding a π phase jump when crossing a field node of the higher order mode). The quadrature condition is thus maintained. A low misalignment induced power imbalance between the output waveguides is thus guaranteed by the design of the interference section.

The devices are single-mode in the vertical axis and thereby do not present an improvement to conventional inverse tapers in this direction. As a result, simulations show a vertical alignment tolerance of only ±0.5 μm. However, in the case of flip chip attachment the alignment accuracy in the vertical direction can be defined by a mechanical contact between the laser and the Silicon Photonics (SiP) chip. For example, vertical alignment can be fine-tuned by the depth of the etch applied to the region of the SiP chip where the laser is attached [11,12]. In this case, the edge defined by the etch process, and not the actual edge of the chip, is also the edge of our edge-coupling device. Attachment methods based on eutectic or thermocompression bonding can achieve vertical alignment accuracies of few hundreds of nanometers [11].

Devices were designed to be compatible with a standard SOI platform offered by ePIXfab consisting of a 2 μm thick buried SiO₂ layer on top of a Si substrate [15]. As depicted in Fig. 3, the waveguides have a Si core with a height of 220 nm and are clad with a 1.25 μm thick SiO₂ top layer.

The rationales for the device design are systematically described in the following: First, modal analysis simulations of the input coupling cross section were performed in order to determine the parameters that maximize the coupling efficiency into the first two modes of the coupling device (calculated at the interface to the chip) within a targeted misalignment range between laser diode and edge coupling device. Then, the length and shape of the interference and splitting section were adjusted to achieve an equal splitting of power between the two output waveguides.

### B. Laser Beam Profile

The different devices have been designed to couple light emitted in the optical C-band (1530-1565 nm wavelength) by a commercial Fabry-Pérot laser diode. The laser emits with linear transverse-electric (TE) polarization. The field profile has been modeled as a Gaussian beam with a typical 1/e half width ($\sigma_x$) of 1.23 μm and 1/e half height ($\sigma_y$) of 1.1 μm at its waist at the facet of the laser (z=0) and increasing monotonically as it propagates towards the chip (along the z-axis), following the equation:

$$\sigma(z) = \sigma_0 \cdot \sqrt{1 + \left(\frac{z}{z_0}\right)^2} \qquad (1)$$

where $z_0$ is the Rayleigh length given by $z_0 = \dfrac{\pi \cdot \sigma^2}{\lambda}$. Within $z_0$, the beam only experiences a slight divergence with $\sigma(z_0) = \sqrt{2} \cdot \sigma_0$. In our case, the laser diode features a $z_0$ larger than 2.4 μm.

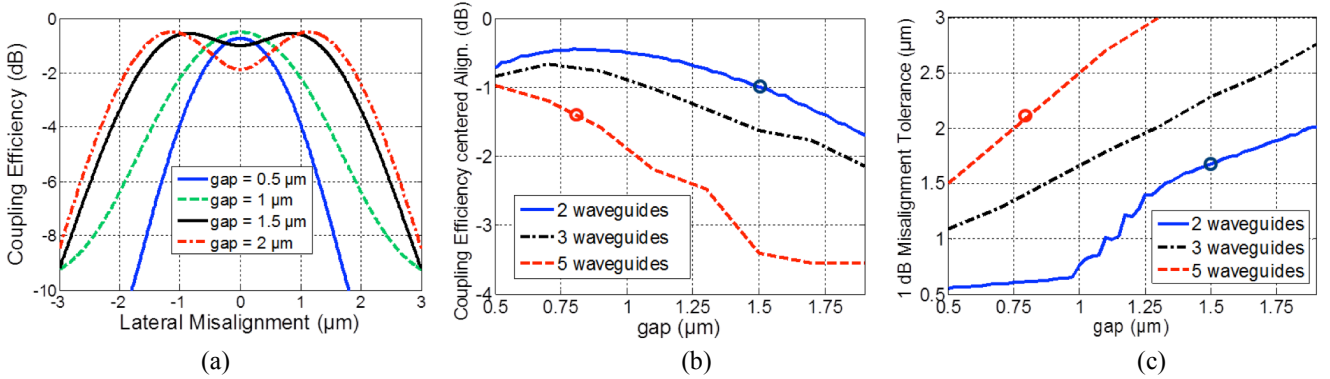

Fig. 4. Trade-off between horizontal misalignment tolerance and coupling efficiency (simulation results): (a) Power coupled into the first two modes of the type I edge coupler (2 waveguides at the input) as a function of misalignment and gap size (b) Power coupled into the first two modes as a function of the gap size and of the number of waveguides assuming centered alignment (c) 1 dB Misalignment tolerance as a function of the gap size and number of waveguides, defined as the lateral misalignment at which the power coupled into the first two modes drops by 1 dB relative to the centered alignment. The design points retained for our two devices have been marked with circles.

Additionally, the free-space propagation of the laser beam also produces a curvature in the wavefront, with a radius given by the equation

$$R(z) = z \left[ 1 + \left( \frac{z}{z_0} \right)^2 \right] \qquad (2)$$

As a consequence of the divergence of the laser beam, the distance between the laser diode and the facet of the chip influences the edge couplers' performance and limits the misalignment tolerance in the z-direction: First, a larger distance between the two elements results in an expansion of the laser beam and additional losses due to mode mismatch in the vertical direction. Second, the wavefront curvature leads to a phase shift between the excited ground and first order modes. This phase difference between the modes at the input of the coupler increases with the distance between the chip and the laser diode. While a target distance can be accommodated by adjusting the length of the interference sections, a deviation from the target results in a departure from the quadrature interference condition described previously and thus in an imbalance at the device output penalizing its worst case performance.

Some flip-chip strategies can be applied to reduce the problem of the misalignment along the z-axis, such as making use of the edge of the chip as a mechanical stopper during the bonding process and taking advantage of the surface tension of the liquid solder [13].

### C. Input Coupling Section

In order to achieve a good mode matching in the vertical direction, similarly to single-mode inverse tapers, the field profile of the modes needs to be delocalized outside the waveguide core by decreasing the tip width. At the same time, the width of the edge coupling device can be increased with the inclusion of additional waveguide tips routed in parallel. Moreover, by adjusting the number of waveguides, their width and spacing (gap), it is possible to control the number of guided modes supported by the super-structure and to tailor their field profile.

The simulation tool FIMMWAVE [16] has been used to obtain the field distribution of the TE modes as a function of the aforementioned parameters ($\lambda_0$ = 1550 nm). Then, the coupling efficiency to the first two modes of the input section has been evaluated as a function of the horizontal and vertical input misalignment (full-vectorial overlap integral). The laser beam was assumed to be flush with the facet of the chip and the losses due to the reflection at the facet were not included at this point.

A waveguide width of 180 nm has been found to be near optimum to match the laser beam profile in the vertical direction. On the other hand, the gap between waveguide tips has a higher influence on the number of guided modes and their profile in the lateral direction. For large gaps, the waveguides are uncoupled and the number of modes equals the number of waveguides. Conversely, a reduction of the gap reduces the number of guided modes.

Fig. 4(a) illustrates the effect of the lateral misalignment on couplers with only two waveguides separated by varying gap values. For gaps smaller than 1 μm, the structure is single-mode and the lateral misalignment tolerance is given by the field profile of the ground mode (± 0.6 μm lateral offset introduces 1 dB loss). On the other extreme, an excessive waveguide separation of 2 μm results in a reduction of the coupling efficiency in the case of perfect centered alignment since the laser is emitting the input beam right in between the two highly separated waveguides. This trade-off between misalignment tolerance and perfectly centered coupling efficiency is depicted in Figs. 4(b) and 4(c).

It is possible to further increase the misalignment tolerance by adding additional input waveguides at the input cross-section of the edge coupler. However, the width of the input cross-section of the edge coupling device is ultimately limited by the laser beam profile: Too wide of an input cross section results in the excitation of higher order modes, which, as we will describe in the following, is not conducive to our designs. Figs. 4(b) and Fig. 4(c) also depict the 1 dB lateral misalignment tolerance and perfectly centered coupling efficiency, both defined as the metrics for the power coupled to the first two modes, for couplers with respectively 3 and 5 input waveguides.

Two different design points have been selected: first, an input coupler with only two input waveguides (type I), a width of 180 nm and a gap of 1.5 μm. Based on the modal analysis, the expected 1 dB loss misalignment tolerance is ±1.65 μm and the insertion loss for centered alignment is at least 1 dB as limited by mode mismatch.

The interface of the second device (type II) is comprised of five 180 nm wide waveguides separated by a 0.8 μm gap. A 1 dB lateral misalignment tolerance of ± 2.2 μm and insertion losses of at least 1.4 dB are predicted based on modal overlaps at the input interface. Two different approaches have been selected for the interference and splitting sections.

### D. Edge coupler type I (two input waveguides)

In the first coupler design (type I) the two input waveguides have been prolonged for a certain length with a constant width (180 nm) and gap (1.5 μm) in the interference section. Then, the waveguides are tapered up to the interconnection waveguide width (400 nm). Finally, the separation between the waveguides is increased with one S-bend on each arm in order to optically isolate them (Fig. 1(a)). In the ideal case of a perfectly aligned laser, only the ground mode (symmetric mode) is excited at the input and a balanced distribution of power is achieved at the output ports independently of the coupler length. On the other hand, a misalignment of the laser produces an excitation of both symmetric and anti-symmetric modes. Due to their different propagation constants, the power couples back and forth between the two waveguides along the interference section. Thus, in order to assure an equal distribution of power between the output ports, the length of the coupler must be designed and fabricated with precision to achieve an accumulated phase difference of $\pi/2$ between super-modes. In the design process, first the required lengths of tapers and S-bends were determined with 3D Finite-Differences Time-Domain (FDTD) simulations to exhibit negligible losses when they are excited with either mode (less than 0.1 dB). Then, based on modal analysis, the length of the interference section (L) was fixed to a quarter of the beat-length ($L_\pi/2$) at the center wavelength of interest ($\lambda_0=1550$ nm), following the equation:

$$L = \frac{L_\pi(\lambda_0)(1+2m)}{2} = \frac{\lambda_0(1+2m)}{4(n_{eff0}(\lambda_0) - n_{eff1}(\lambda_0))} \quad (3)$$

with $n_{eff0}$ and $n_{eff1}$ respectively the effective indices of the ground and first order mode and $m$ an integer. The additional term in $2m$ corresponds to adding an integer number of $\pi$ phase shifts and should be set to zero in the above. An $L_\pi/2$ of 22.8 μm has been obtained in simulations.

Finally, 3D FDTD simulations of the whole structure including the laser beam modeled in an external air region were performed to fine tune the interference section length, also taking into account the coupling in the tapered and S-bend sections. A possible deterministic misalignment in the z-axis, as given by the assembly process, should be considered at this point of the design flow. A total insertion loss of 1.42 dB in the case of centered alignment was obtained for an optimized interference section length (L) of 16 μm.

The beat-length of the interference section depends on the wavelength and consequently limits the bandwidth of the device. A wavelength of operation differing from the design wavelength results in the quadrature condition not being perfectly satisfied and increases the power imbalance at the output waveguides. Simulation results show a 1 dB bandwidth of 90 nm for the designed device, defined as the wavelength range that produces a 1 dB of additional loss in the worst output waveguide.

### E. Edge coupler type II (with five input waveguides)

The second coupler (type II) is comprised of 5 parallel input waveguides, 180 nm wide and separated by 0.8 μm gaps. The complexity of this device is higher than the previous one, not only due to the higher number of waveguides, but also due to the presence of an additional guided second order mode at the input. The excitation of the second order mode is detrimental and a mechanism has to be provided to cut off and extinguish this mode: Even though the second order mode can increase the power coupled to the two output waveguides under certain conditions, when the amplitude of the mode is inverted due to a displacement of the laser beam (crossing a node of the mode, at a laser misalignment larger than ± 1 μm, see Fig. 2), a destructive interference will occur worsening the worst case coupling efficiency of the device. The previous technique of using a quadrature phase shift to prevent interference cannot be applied, as the mode cannot be in quadrature to both the ground mode and the first order mode simultaneously. Hence, the second order mode should be extinguished if a misalignment tolerance larger than ±1 μm is desired.

The coupling device is comprised of five sections indicated in Fig. 1(b) as (1) to (5):

The first part consists of a prolongation of the input waveguides. In simulations, the input section profile presents a half beat length ($L_\pi$) of 54.2 μm at $\lambda_0$. In order to avoid an unnecessary limitation in the bandwidth derived from the accumulated phase shift between the ground and the first order modes, the length of this section should be kept as short as possible. In this study, its length was limited by the precision of the dicing saw. By defining the input interface of the edge coupler with a lithographically defined etch, this length could however be controlled within a few tens of nanometers and be drastically reduced.

Next, the five waveguides are tapered up to a width of 0.88 μm in the second section and combined into a single 5 μm wide rib waveguide. This waveguide has been defined with a shallow etch of 70 nm. The final gap between the waveguides was reduced down to 100 nm prior to merging, as limited by the critical dimension of the process. The transition from full to shallow etch waveguides introduces 0.8 dB of additional loss. The length of the tapers was set to 50 μm to adiabatically convert the modes excited at the input, avoiding cross-coupling to higher order modes.

In the next section (3), the rib waveguide is again tapered down to 1 μm, below the cut-off of the second order mode so as to filter it out. A taper length of 50 μm was chosen to convert the spot size of the ground and first order modes with negligible losses (below 0.1 dB).

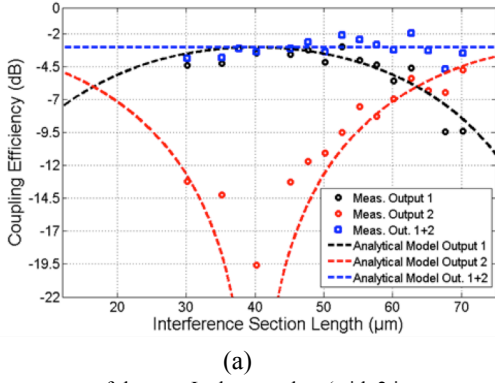

(a)

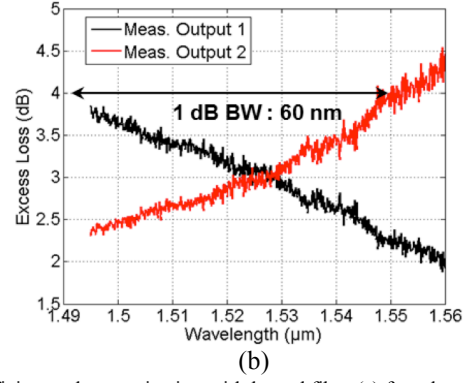

(b)

Fig. 5. Measurements of the type I edge couplers (with 2 input waveguides): Coupling efficiency characterization with lensed fiber (a) for edge couplers with different interference section lengths (b) as a function of the wavelength. Dots in (a) represent measurement results and dashed curves modeling results. The curves in (b) are measured.

Next, in section 4, a short rib waveguide was inserted in order to adjust the phase shift between the ground and first order modes. The half beat-length between the two lowest modes of this section is 4.65 μm at $\lambda_0$.

The light is then split into two 0.45 μm wide rib waveguides with a Y-junction (100 nm initial gap between output waveguides). The rib profile of the waveguides allows a power splitting with negligible losses for the ground and first order modes (less than 0.1 dB).

Finally, adiabatic waveguide transitions between the rib and the fully etched interconnection waveguides are included (section (5)). The total length of the coupler is 156.7 μm.

3D FDTD simulations of the complete device show insertion losses of 2.2 dB when the structure is excited with the modeled laser beam aligned at the central position in an external air region. The length of the input section and the central interference section (L) were fixed to 5 μm and 1.7 μm respectively to obtain the quadrature interference condition and an equal splitting of power between the outputs.

The lengths of the different tapers in the interference section have been designed to safely guaranty adiabatic transitions with low insertion losses. However, an excessive length reduces the bandwidth. The quadrature condition is obtained if the accumulated phase difference is an odd multiple of $\pi/2$, i.e., $\Delta\varphi = (2 \cdot m + 1) \cdot \pi / 2$ for any integer $m$. However, the bandwidth of the device ($\Delta\lambda$) is inversely proportional to the total phase shift: Ignoring waveguide dispersion, $\Delta\lambda / \lambda_0$ scales as $1/(2m+1)$. For the designed device, the total phase difference is $9\pi/2$ and the bandwidth is limited to 35 nm. If a higher bandwidth is required, a central taper (section 3) with a length reduced to 30 μm is preferred. This reduction results in a total phase difference of $7\pi/2$ without compromising the insertion losses excessively (0.2 dB of additional loss). In this case, the bandwidth is increased to 50 nm as given by resimulating the complete device.

## III. EXPERIMENTAL RESULTS

### A. Fabrication

The devices have been fabricated within the ePIXfab Network in the 200 mm wafer IMEC CMOS pilot line with 193nm Deep UV lithography and dry etching [15]. The different structures were defined on a SOI wafer with a buried oxide layer thickness of 2 μm and a silicon core layer thickness of 220 nm. The edge couplers were placed near the dicing lanes of the chip, the facets of which were later hand polished. For the sake of a higher density of integration and simplicity in the testing process, the two outputs of the edge couplers were connected to individual grating couplers. The fabrication of these grating couplers requires a shallow etch step (70 nm depth) that is part of the standard photonic IMEC process. This shallow etch was also used in the definition of the rib waveguides inside the type II edge couplers (Fig. 1(b)).

After dicing, an edge polishing process was necessary to access the edge-coupler interface and provide a smooth facet. To compensate for the limited precision of the hand-polishing process and in order to investigate the effect on imbalance and reproducibility of the proposed devices, edge coupling structures with different interference section lengths were included on the chip for both types of devices.

The grating couplers enable fast and reliable alignment of the edge coupler outputs with a fiber array by means of an automatic alignment stage. As a drawback, the gratings introduce additional losses that needed to be accurately estimated and subtracted to extract the insertion losses of the proposed edge couplers. For this purpose, 39 grating coupler loops were also included on the chip. The measured variability of the grating coupler coupling efficiency introduces an uncertainty of ± 0.3 dB (standard deviation) in the determination of the edge coupler insertion losses.

### B. Optical Characterization with tunable laser and lensed fiber

After the alignment of the output grating couplers, the edge couplers were characterized with a tunable laser connected to a lensed fiber. The lensed fiber provided a light source profile with similar dimensions (2.5 μm spot size, 1/e half width and half height σ = 1.25 μm) to the Fabry-Pérot laser mode and allowed a reliable characterization over the entire bandwidth of interest. Moreover, the lensed fiber allowed a measurement of the reflection induced by the edge-coupler back into the single mode lensed fiber (the relevant metric to also assess the reflection back into a single mode laser diode of similar spot size, i.e., the overlap integral of the optical reflection with the laser mode profile is already taken into account). Finally, the focal distance of the lensed fiber (14 μm) facilitated measurements by providing a safeguard and preventing damage to both elements due to an abrupt collision during characterization.

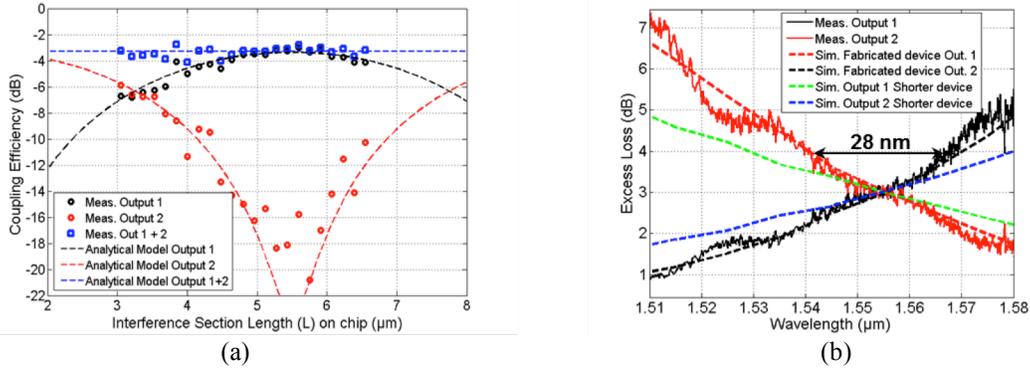

(a)  (b)

Fig. 6. Measurements of type 2 edge couplers (with 5 input waveguides): Coupling efficiency characterization with lensed fiber (a) for edge couplers with different rib waveguide lengths (section (4) in Fig. 1(b)) and (b) characterization of the device that is balanced at 1550 nm as a function of the wavelength. Dots in (a) represent measurements and dashed curves modeling results. The continuous curves in (b) are measured, the blue and red curves correspond to modeling of the measured device and the blue and green curves show modeling of the higher bandwidth device with shortened MMI section (section 3 in Fig. 1).

In order to measure the coupling efficiency as a function of the input alignment, the lensed fiber was mounted on an automatic XYZ nanopositioner stage. For each coupling structure, the power at both grating coupled outputs was monitored as a function of the position of the fiber. During the measurements the polarization was fixed to linear TE.

First, the type I edge couplers (two input waveguides) were measured at a wavelength of 1550 nm. Fig. 5(a) shows the measured output power for edge couplers with different interference section length (L) after normalizing out the losses due to the grating couplers. The fiber alignment in Fig. 5 corresponds to the position that produces the maximum coupling efficiency but also the maximum imbalance (lateral misalignment of 0.9 μm, see also Fig. 4(a)). This alignment is chosen here as the goal of this set of measurements is to characterize the imbalance (note that the insertion losses are defined at the center alignment in this paper, where the coupling is slightly worse). The half beat-length derived from the measurements (~23.5 μm) agrees well with the value obtained in simulations (22.8 μm). The device with the lowest imbalance between the output ports achieves the quadrature condition with a total phase difference of 3π/2 between the super-modes. The best device (62.7 μm length) exhibits excess insertion losses of 1.8 dB for an alignment at the central position (close to the 1.42 dB calculated in simulations). The scatter in the measured loss data is attributed to local defects on the chip facet caused by the dicing saw that were not fully removed by polishing (this defectivity was confirmed by microscope inspection of the facet).

The imbalance between output ports as a function of wavelength has been also characterized. Fig. 5(b) shows the excess loss of each output of the device that is balanced at 1550 nm (with 60.7 μm interference section length). The device exhibits a 1 dB loss bandwidth of 60 nm. Simulations results showed that this value could be increased above 90 nm by reducing the length of the interference section to 16 μm (π/2 phase difference between the ground and first order modes). Experimentally, we were limited by the precision of the dicing saw we used. However this limitation would not apply to a lithographically defined device edge.

Next, the type II edge couplers (five input waveguides) were characterized at 1550 nm. The devices differ in the length of the central interference section (No. 4 in Fig. 1(b)). The initial gap of the Y-junction turned out to be slightly below the process resolution and the fabricated devices present an additional interference length of 2.6 μm. Additionally, the length of the input section after dicing and polishing was 29 μm. Both are compensated for by an additional length in section 4 in order to maintain quadrature (L=3.3 μm instead of 1.7 μm as initially simulated). The measurement results are plotted in Fig. 6(a). The best device exhibits a total coupling efficiency of -2.8 dB with a centered alignment (close to the -2.2 dB predicted by simulation). In order to obtain reliable device fabrication, the initial gap of the Y-junction should be chosen sufficiently large to be deterministically fabricated, since this device is sensitive to phase propagation (as opposed to a classic Y-junction).

Fig. 6(b) shows the measured bandwidth of the edge coupler with 5 input waveguides. Due to the larger phase shift (11π/2, larger m), the measured bandwidth, 28 nm, was smaller than initially predicted. Simulation results show that by reducing the length of the input section (No. 1) to 5 μm and the taper length in section 3 to 30 μm, the bandwidth can be increased to 50 nm (phase shift 7π/2).

Finally the reflection level was measured by means of a fiber optic circulator connected to the lensed fiber. In the experiment, neither the lensed fiber nor the chip facet was covered with an antireflection coating. As both the lensed fiber and the laser diode were shown to have beam profiles of similar dimensions, the power that is reflected in the facet and coupled back into the fiber is a good metric to evaluate the level of feedback that can affect the performance of the laser. The output of the edge coupling device and the back-reflection into the lensed fiber were jointly monitored as a function of fiber misalignment. Losses due to the circulator were normalized out from the reflection data. Fig. 7 shows the obtained results for the balanced structure of type II (five input waveguides). In the case of perfect vertical alignment (Fig. 7(a)), the device exhibits a reflected power of less than -20 dB, as is primarily given by the $SiO_2$-air interface due to the tiny fill-factor of the silicon waveguide tips (0.18 μm). When applying vertical misalignment (worst case plotted in Fig. 7(b)) the level of reflection increases drastically in the direction of the substrate once the beam sees the substrate due to the high silicon-air index contrast. However, inside the 1 dB vertical misalignment tolerance of the device (±0.5 μm), the variation in the reflected power is negligible.

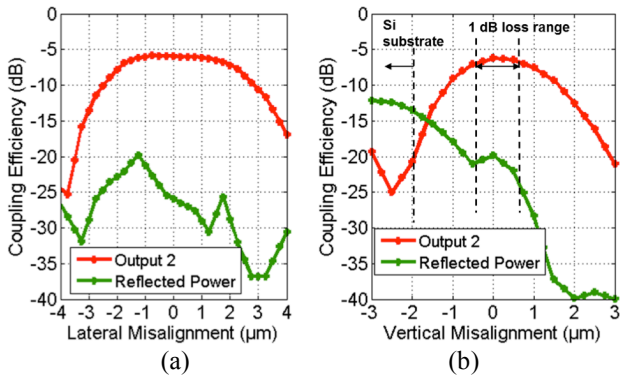

(a)                                    (b)

Fig. 7. Reflected power of type II edge coupler (five input waveguides) and coupling efficiency to edge coupler output 2 as a function of the (a) lateral misalignment (for perfect vertical alignment) and (b) vertical misalignment (worst case positioning along the x-axis corresponding to maximized back-reflection).

### C. Coupling efficiencies and Misalignment tolerances with Fabry-Pérot laser.

In the final set of measurements, the best structures in terms of imbalance were characterized with a commercial InGaAsP Fabry-Pérot laser diode from Archcom with a die size of 250 μm x 350 μm [17]. The laser was wire-bonded onto a standard submount for electrical connection. The submount was additionally welded to a Peltier element for temperature control and to an aluminum plate for mechanical stability. Finally, in order to measure the coupling efficiency as a function of the alignment position, the plate was mounted on the automatic nanopositioner stage. During the measurement process, the laser was biased with 25 mA and the temperature controller was set to 25° C. At this bias point the total output power of the laser was 2.04 dBm. After alignment, the spectrum of the laser was measured with an OSA connected at one of the outputs. At this bias point, the laser was emitting at 1537 nm with a side lobe suppression ratio of -27 dB.

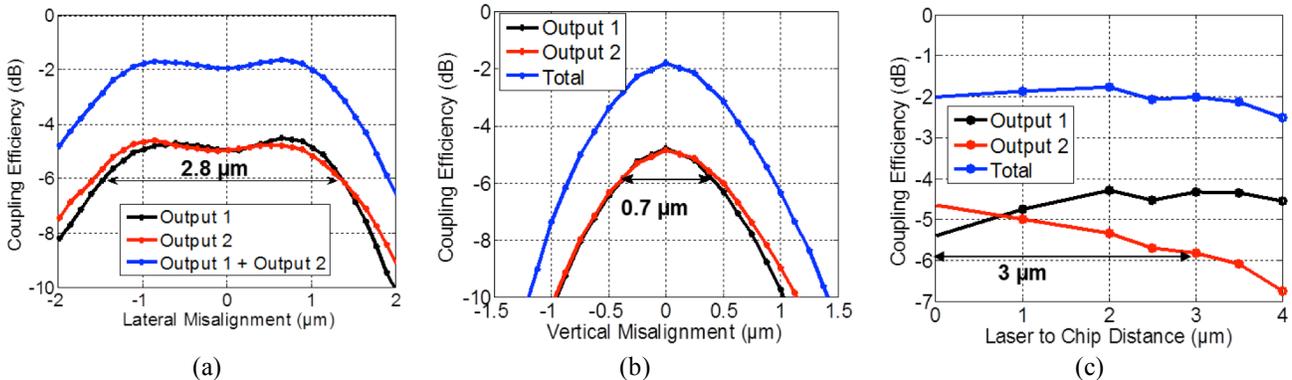

(a)                                    (b)                                    (c)

Fig. 8. Misalignment tolerance and excess insertion loss characterization with Fabry-Pérot laser diode for the type I edge coupler (with two input waveguides): (a) Horizontal Misalignment (b) Vertical Misalignment (c) Laser to chip distance tolerance.

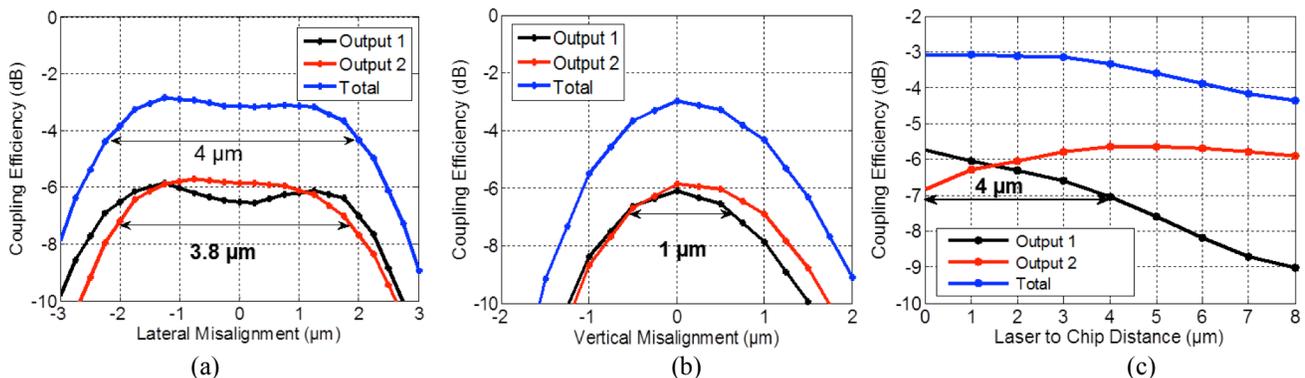

(a)                                    (b)                                    (c)

Fig. 9. Misalignment tolerance and excess insertion loss characterization with Fabry-Pérot laser diode for the type II edge coupler (with five input waveguides): (a) Horizontal Misalignment (b) Vertical Misalignment (c) Laser to chip distance tolerance.

Figs. 8 and 9 show the misalignment tolerances measured in the three axes after normalizing to the input power and subtracting the grating coupler insertion loss at the emission wavelength. The edge coupler with two input waveguides exhibits a horizontal misalignment tolerance of ±1.4 μm and a total insertion loss at the central alignment position of 2 dB (close to the 1.8 dB measured with the lensed fiber and the 1.4 dB predicted by simulation). As expected from simulations, the edge coupler with five input waveguides performs better in terms of horizontal misalignment tolerance (±1.9 μm) but has a lower total coupling efficiency at the central position (3.1 dB insertion loss, close to the 2.8 dB measured with a lensed fiber and the 2.2 dB predicted by simulation). In relation to the vertical alignment tolerance, both devices present sub-micrometric tolerances: ±0.35 μm and ±0.5 μm, respectively for the cases of two (type I) and five (type II) input waveguides. Furthermore, the device with five input waveguides (type II) has been found to have a better tolerance to laser alignment in the z-axis (± 2 μm) than the type I edge coupler with 2 input tips (±1.5 μm). These z-axis alignment tolerances are consistent with simulations and are related to the fact that the accumulated phase difference between the super-modes at the input of the edge coupler depends on the distance between laser and chip relative to the beat length of an equivalent air-waveguide with the same width as the input section of the edge-coupler. This beat length is longer for a wider input section, relaxing the z-axis tolerance of the type II device (5 input waveguides). Table I summarizes the measured characteristics of the two devices.

TABLE I
COMPARISON OF THE TWO DEVICE TYPES

|  | Type I | Type II |
|---|---|---|
| Excess Insertion Loss[1] | 2 dB | 3.1 dB |
| x-z 1 dB tolerance[1] | ±1.4 μm | ±1.9 μm |
| y-axis 1 dB tolerance[1] | ±0.35 μm | ±0.5 μm |
| Measured Bandwidth[2] | 60 nm | 28 nm |
| Improved Bandwidth[3] | 90 nm | 50 nm |

[1]Measured with laser diode
[2]Measured with tunable laser
[3]Simulated assuming lithographic control of device edge

## IV. CONCLUSIONS

Two different edge couplers that relax the lateral alignment accuracies required in hybrid integration of laser diodes with sub-micrometric SOI waveguides have been demonstrated. The presented devices overcome the trade-off between peak coupling efficiency and lateral misalignment tolerance existing in conventional edge couplers by splitting the injected power between two single mode waveguides at the device output with almost equal power distribution but with a varying relative phase that accommodates the lateral misalignment of the laser and the coupling structure. The devices have been experimentally characterized with a lensed fiber and a commercial Fabry-Pérot laser diode. The two presented devices exhibit an excess insertion loss of respectively 2 dB and 3.1 dB (worst case in either waveguide). Furthermore, an outstanding 1 dB misalignment tolerance of respectively ±1.4 μm and ±1.9 μm has been measured without reshaping the beam of a commercial Fabry-Pérot laser. Comparable 1 dB tolerances have been obtained for the laser to chip distance (±1.5 μm and ±2 μm ranges respectively). A very low reflection level from the device edge of less than -20 dB has been measured within the 1 dB alignment range.

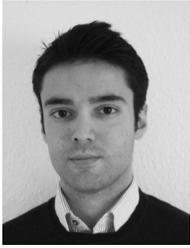 **Sebastian Romero-García** was born in Rute, Spain in 1984. He received the M.Sc. in telecommunications engineering from the University of Malaga, Spain, in 2009. Since 2011, he has been with the Integrated Photonics Laboratory of RWTH Aachen University where he is currently working towards the Ph.D. degree in electrical engineering. From 2009 to 2011, he was a Research Assistant with the Department of Communications Engineering, University of Málaga. His research interests focus on the development of integrated photonics devices and systems.

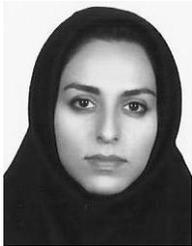 **Bahareh Marzban** was born in Shiraz, Iran in 1988. She received her B.Sc. in Electrical Engineering from Shiraz University Iran in 2010. Since 2011, she has been working towards a M.Sc. degree in Communication Engineering at RWTH Aachen and in 2012 she joined the Integrated Photonics Laboratory where she wrote her Master thesis. Currently she is doing an internship in partial fulfillment of the requirements for her Master's degree at Innolume GmbH.

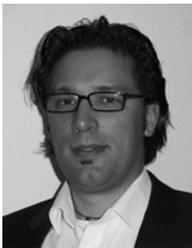 **Florian Merget** was born in Bochum, Germany in 1977. He received his Dr.-Ing. degree in electrical engineering from RWTH Aachen University in 2008. From 2008 to 2011 he was research assistant with the Institute of Semiconductor Electronics of RWTH Aachen. In 2011 he joined the newly founded Integrated Photonics Laboratory as staff scientist. His research interest includes high-speed modulation devices and high datarate optical communication systems that are specifically targeted to meet the needs of next generation telecom and datacom applications.

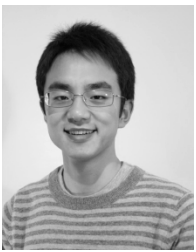 **Bin Shen** was born in Zhejiang, China in 1986. He received his M.Sc. in Materials Science form RWTH Aachen University in 2012. Since 2011 joined the Integrated Photonics Laboratory (IPH). Now he is a Ph.D. (Dr.-Ing.) candidate at IPH.

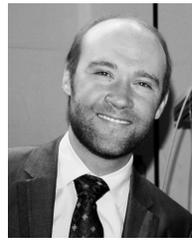 **Jeremy Witzens** was born in Schiltigheim, France in 1978. He received his Engineering Diploma from the Ecole Polytechnique (Palaiseau, France) in 2000 and his Ph.D. from the California Institute of Technology (Pasadena, USA) in 2005. He has been a Professor at RWTH Aachen University and the head of the Integrated Photonics Laboratory since 2011. From 2009 to 2010 he was a Principal Research Scientist at the University of Washington (Seattle) and from 2006 to 2009 a Sr. Staff Engineer at Luxtera Inc. The Integrated Photonics Laboratory is currently focusing on Silicon Photonics devices and systems, as well as novel materials for group IV photonics.